# Comparison of h-BN and graphene layers as grain boundary materials for granular FePt-L1$_0$ thin films

B.S.D.Ch.S. Varaprasad[1, 2], Chengchao Xu[1,2], Brandon Reese[1,3], David E. Laughlin[1,2, 3] and Jian-Gang Zhu[1, 2,3]

1) Data Storage Systems Center, Carnegie Mellon University, Pittsburgh, PA, USA
2) Electrical and Computer Engineering, Carnegie Mellon University, Pittsburgh, PA, USA
3) Materials Science and Engineering Department, Carnegie Mellon University, Pittsburgh, PA, USA

**Abstract**

Granular L1$_0$-FePt thin films with small columnar grains are essential for heat-assisted magnetic recording media. While hexagonal boron nitride(h-BN) has proven effective for promoting columnar FePt grains, we explored multilayer graphene as an alternative grain boundary material leveraging its structural similarity to h-BN. The FePt granular thin films with carbon-based grain boundary materials(GBMs) were deposited by cosputtering on Si/SiO$_2$ substrates with substrate bias at 650°C. The RF bias and high temperature facilitated formation of interlinked graphene nanoribbons wrapping around FePt grains, yielding 7.5 nm diameter, 8 nm height grains with an order parameter of 0.78 and a perpendicular coercivity of 40 kOe. However, the formation of graphene nanoribbons could not effectively promote columnar structures, likely due to co-existing amorphous carbon in grain boundaries. Optimizing deposition to improve graphene grain boundary quality is necessary to realize this 2D material's potential for achieving desirable microstructures for HAMR media.

Corresponding author: sbollapr@andrew.cmu.edu

**Introduction**

For the realization of next generation heat-assisted magnetic recording (HAMR) media, granular film structures should exhibit high grain aspect ratios ($h/D > 1.5$) and good thermal stability up to T ~ 600° C [1]. Currently, ordered $L1_0$-FePt has attracted considerable attention for HAMR media applications due to its high magneto-crystalline anisotropy that exhibits a desirable temperature dependence near its Curie transition [2]. Tall FePt grains in the media films are vital to ensure a high signal to noise ratio (SNR) for HAMR applications [3]. To achieve aspect ratios greater than $h/D = 1.5$ careful consideration must be given when selecting a grain boundary material (GBM). Several studies regarding amorphous GBMs like C [4], $TiO_2$ [5], $SiO_2$ [6, 7] and $Cr_2O_x$ [8] have successfully pushed $L1_0$-FePt grain aspect ratios to around 2 [9, 10]. Despite their effect on promoting columnar growth, all amorphous GBMs exhibit one or more drawbacks, including degradation of FePt ordering, low thermal stability or poor in-plane microstructures. Recently, FePt media with high aspect ratios ($h/D \approx 2.5$) [11] and good thermal stability have been reported in the FePt-boron nitride (BN) system. In this study FePt and BN were co-sputtered, with an additional RF bias applied to the substrates. The substrate bias facilitated the growth of hexagonal-BN (h-BN) nanosheets perpendicular to the film plane wrapping around the FePt grains side surfaces, which led to highly columnar grain structures. Without a proper substrate bias, h-BN cannot be formed; instead, only amorphous-BN (a-BN) is formed in the grain boundaries [12]. These findings suggest that other amorphous GBMs may exhibit similar crystallization behaviors associated with substrate bias during sputter deposition.

The FePt-C system is another system well known to exhibit excellent grain isolation and good thermal stability. However, introduction of amorphous carbon (a-C) at the grain boundaries has not enabled forming tall and columnar FePt grains, typically failing to achieve aspect ratios ($h/D$) greater than one [4]. While a-C struggles to foster columnar grain structures, graphene, one of its crystalline allotropes, is structurally similar to h-BN and can be a good candidate of GBMs. Lattice parameters of graphene and h-BN are $a = b$ = 2.47 / 2.51 Å, $c$ = 6.71 / 6.61 Å respectively. Furthermore, graphene and h-BN exhibit comparable mechanical properties of high in-plane strength but extremely low flexural rigidity [13], similar to other 2D materials. The high bendability makes it possible for graphene layers to wrap around the columnar grains like h-BN nanosheets.

Considering graphene's many desirable qualities [14], we have conducted an experimental study evaluating the effect of RF substrate bias on the growth of FePt-C media. FePt-C media deposited with RF substrate bias have been prepared and relevant magnetic, structural, and transmission electron microscopy (TEM) data are presented below.

**Experimental**

A multilayer (ML) film stack of {Ta (2 nm) | Cr (30 nm) | MgO (10 nm) | FePt (0.4 nm) | [FePt-C$_{30vol.\%}$ (0.5 nm) without RF bias] | [FePt-C$_{30vol.\%}$ (X nm) with RF bias]}, referred to as FePt-C-ML, was deposited on the Si | SiO$_2$ substrates using an ultra-high vacuum AJA sputtering system with $1 \times 10^{-8}$ Torr base pressure. The thickness of the top layer, X = 4~7 nm, varies from sample to sample. Consequently, total thickness of the FePt-C magnetic layers on MgO underlayer are $t$ = 5, 6, 7, 8 nm, respectively. Ta, Cr, MgO, C, Fe$_{55}$Pt$_{45}$ alloy, and BN targets with 99.9% purity are used to deposit films. Ta was deposited at room temperature as an adhesion layer for SiO$_2$ substrate. One minute of pure O$_2$ exposure (0.1 mTorr) was adopted to modify the Ta surface to attain good Cr texture. A 30 nm thick Cr layer was deposited at 250℃ to realize the good (200) texture on Ta. The substrate was then annealed at 600℃ for 1 hr after the Cr deposition to attain large grain sizes in the Cr layer and to reduce the surface roughness. The MgO underlayer was deposited at room temperature with 100W RF-power, 10

mTorr, and the target to substrate distance was around 80mm. To improve the nucleation density, an ultrathin (0.4 nm) layer of pure FePt was deposited prior to the co-sputtering of FePt and carbon, as was discussed in our previous papers. Both RF bias and high-temperature deposition were found to be necessary factors for the formation of graphene layers. A low RF bias (5W, $V_{DC} \approx$ -55V) was applied to the substrate mainly intended to induce the formation of graphene layers. All the samples were sputtered using the same conditions, except for the deposition times of the top FePt-C (with RF bias 5W, $T = 650°C$) which were varied to achieve different film thicknesses. Standard X-ray diffraction (XRD) with a Copper-Kα source was used to analyze the film texture and order parameter. The microstructure of the samples was evaluated by in-plane and out-of-plane transmission electron microscopy (TEM) imaging, using bright-field TEM (BF-TEM), high-resolution TEM (HR-TEM), and scanning TEM-high angle annular dark field (STEM-HAADF) techniques using FEI Titan Themis 200. The grain size and grain center-to-center pitch distances were analyzed using the in-plane STEM-HAADF images and the image processing software (MIPAR). The magnetic moment (*M*) vs. field (*H*) curves of the film samples were measured with a Quantum Design Superconducting quantum interference device-vibrating sample magnetometer (SQUID-VSM).

**Results and Discussion**

As mentioned above, it has been shown that h-BN is a structurally stable GBM at high temperatures which gives good $L1_0$-FePt ordering [10]. Here, we have investigated the use of graphene layers as the GBM for ordered $L1_0$-FePt granular thin films. The XRD patterns for all FePt-C-ML samples deposited at $T = 650°C$, shown in Fig. 1, reveal strong FePt (001) texture and good $L1_0$ ordering. The intensity ratios ($I_{001}/I_{002}$) range from 2.2 for the 5-nm sample to 2.4 for the 8-nm sample, indicating good chemical ordering. Specifically, the FePt (002) peak position near $2\theta \approx 48.7°$ for all samples is consistent with previous reports on $L1_0$-FePt thin films [4, 9, and 10]. Additionally, the order parameter values ($S$) have been calculated and are labelled in Fig. 1. Our calculations follow the method presented by Yang et al.: XRD geometry, FePt texture, and film thickness ($t$) have been considered as the correction factors [15]. Small FePt (111) and (200) traces are also observed for the samples thicker than 6 nm, evidenced by the small humps near $2\theta = 41°$ and $47°$ respectively in Fig. 1.

Fig. 2 shows the coercive field ($H_C$) as a function of film thickness. All samples show high perpendicular coercivities, which increase monotonically with increased thickness. The maximum occurs for the 8-nm sample at $H_C = 40$ kOe. The in-plane coercivities also follow an increasing trend as the film thickness increases, indicating an increasing number of misaligned grains with increased film thickness. The highest in-plane coercivities, $H_{C//} \approx 10$ kOe, can be correlated with the emergence of FePt (111) peak observed in the XRD spectra of the 8-nm sample. From these measurements it is estimated that all the samples exhibit strong uniaxial anisotropy with $H_K \approx 70$ kOe and high perpendicular $H_C$'s on the order of 33 to 40 kOe. This is in good agreement with the XRD data which suggests strong $L1_0$ ordering for all samples, $0.76 < S < 0.78$. However, the growing number of misaligned FePt (111) and (200) grains contributes significantly to the observed in-plane coercivity of the samples.

The microstructures of the 5-nm and 8-nm FePt-C-ML samples are presented in Fig. 3. As illustrated in the in-plane HRTEM micrographs, these samples exhibit small spherical FePt grains that are well-isolated by carbon GBMs. The lattice fringes within the grain boundaries indicate the formation of the graphene layers. These graphene layers bend and surround the FePt grains, often forming a stack of 3~6 parallel layers sandwiched by two neighboring grains. Majority of the graphene layers are perpendicular to the film plane. It is also common to observe Y-shaped junctions formed by these few-layer graphene nanosheets (as shown in Fig.3.(b) and (c)). The junction formation is likely resulted from (i) meeting of the growing edges with existing basal planes or (ii) the branching out of the layer-by-layer grown basal planes. Moser J. et al. have demonstrated a similar branching behavior of $MoS_2$ layer growth, and they attributed this growth mechanism to the adsorption of contaminants [16]. As a result, the graphitic GBMs can be considered as a planar network made of interlinked multilayer-graphene nanoribbons, and the FePt grains are embedded in this network matrix.

To gain further insights, additional investigations were conducted through fast Fourier transform (FFT) analysis of the micrographs. The FFT pattern of Fig.3(b) is presented in Fig.3(a), which shows diffraction spots of MgO(200), FePt(200), $L1_0$-FePt(110), and a relatively fuzzy diffraction ring spanning from 2.5~2.8 $nm^{-1}$ (0.36~0.40 nm). The two brighter sections on the ring stem from the (100) planes ($d = 0.371$ nm) of the in-plane ordered $L1_0$-FePt grains, which are labeled by red circles in Fig.3 (b). Moreover, the whole ring primarily comes from the curved graphene layers in the grain boundary area. The interlayer distances of the graphitic multilayers (0.36~0.40 nm) observed in the granular film are therefore larger than the (002) d-spacing of the hexagonal graphite ($d = 0.335$ nm) [17]. Two intensity profiles of the TEM image are measured along the two yellow arrows in Fig. 3(d), through the (001) planes of a FePt grain with its c-axis lying in the plane, and in Fig. 3(e), normal to the graphene fringes sandwiched by two FePt grains. The measurements depicted in Fig. 3(f) reveal that the average interlayer distance of these graphitic multilayers is 0.37 nm. The widening of interlayer distances can also be attributed to the misalignment between the monolayer lattices, likely due to the bending of graphene layers. Similar variations in interlayer distances have been observed in graphene nanowalls[17], folded graphene nanoribbons [18] and twisted bilayer graphene structures [19]. Besides, it has been reported that magnetron sputtering can introduce relatively high point defect concentrations in graphene, likely due to the energetic ion bombardment during the deposition process [20]. Lattice defects can lead to some 3D features, such as the distortion of monolayers and some local $sp^3$ bondings between graphene layers, which widen the graphene interlayer distances. Sufficiently high defect density can form the so-called turbostratic graphite (TG) phase [17]. In summary, curving of the graphene nanosheets in the FePt-graphene granular thin film, and possibly impurity or point defects in the graphene layers contribute to the misalignment between basal planes. This results in the branching growth or some turbostratic characteristics with larger interlayer distance.

Moreover, the contrast between Fig. 3(b) and Fig. 3(c) clearly illustrates the strong tendency of lateral growth of FePt grains with graphene in grain boundaries. The lateral growth of FePt grains is a particle coarsening process with incoming flux of atoms. As described in the Lifshitz-Slyozov-Wagner theory, larger particles are more energetically favored over smaller particles. Consequently, larger grains have a stronger ability to capture atoms, whether from incoming flux or detached from smaller grains, that are absorbed and moving on the film-growing surface. Thin films of FePt with amorphous carbon [4] commonly produce spherical FePt grains due to the high interfacial energy. The high interfacial energy and strong phase separation tendency between FePt and a-C indicates a reluctance to form bonds, which can contribute to a lower diffusion barrier and thus high diffusion rates of one material on the surface of the other. This characteristic of the FePt-(a-C) system may be a root cause that makes it challenging to inhibit the coarsening of spherical FePt particles in the FePt-(a-C) thin films.

As depicted in Fig. 4, grain sizes of the FePt-graphene (30 vol%) samples increase from D =7.5±2.8 nm for $t$ = 8nm to D = 5.5±1.8 nm for $t$ = 5nm. The FePt grain sizes in the FePt-(h-BN) $_{20vol\%}$ samples are also presented as a dotted line with star marks in Fig. 4 for comparison. The plot of grain diameter vs. thickness for FePt-(h-BN) $_{20vol\%}$ samples demonstrates that the introduction of h-BN nanosheets can effectively suppress the lateral growth, whereas FePt-graphene$_{30vol\%}$ samples exhibit a pronounced tendency for lateral grain growth [12].

The out-of-plane TEM micrographs reveal that FePt-graphene samples with thickness below 8nm are single-layered FePt grains. However, for thicker samples, aside from the large FePt grains sitting on the MgO surface, a second layer of smaller FePt grains begins to appear on the top, which are also observable as brighter spots in the in-plane STEM-HAADF images. In contrast, the cross-section HRTEM image of the 12-nm-thick FePt-(h-BN) thin film sample Fig. 5(a) shows a single layer of columnar grains with aspect ratio more than 2. We can see the h-BN nanosheets grow along the FePt grain sidewalls and parallel to the film growing direction. In the case of carbon GBM, as shown in Fig. 5(b), although graphene layers have grown around spherical FePt grains, they coat over the grains during the deposition, leading to the formation of the second layer.

The formation of graphene layers did not significantly differ from previously reported a-C GBMs. The FePt-graphene thin films exhibit similar spherical particle shapes of FePt as observed with a-C. In comparison to FePt-(h-BN) samples, the presence of graphene layers fails to promote the columnar growth of FePt grains or to suppress the lateral grain growth. From another perspective, the observed lateral growth and second layer formation phenomena of FePt grains could suggest that the carbon GBMs deposited in these samples consists of both a-C and graphene. Further optimization of deposition parameters, such as substrate bias voltage and temperature, may be required to improve the quality and continuity of the graphene GBMs. A higher graphene fraction in the GBM could potentially promote the columnar growth of FePt grains and suppress lateral grain coarsening, similar to the behavior observed in the FePt-h-BN system

**Summary**

In this study, we have developed granular $L1_0$-FePt thin films with graphene-based grain boundary materials (GBMs) for potential HAMR media applications. The RF substrate bias and high deposition temperature facilitated the formation of interlinked graphene nanoribbons surrounding the FePt grains. Well-isolated $L1_0$-FePt grains with an average diameter of 7.5 nm and height of 8 nm were achieved in the FePt-graphene-ML films. These films exhibited a high order parameter of 0.78 and a perpendicular coercivity of 40 kOe.

While the graphene nanoribbons formed and wrapped around the FePt grains, they were unable to effectively suppress lateral grain growth or promote columnar grain structures, in contrast to the behavior observed with h-BN nanosheets. Further optimization of the deposition conditions is necessary to improve the quality and continuity of the graphene layers within the grain boundary regions. In the case of co-sputtering a high corban deposition rates are required to match the FePt deposition. Increased C deposition rate in the deposited media leads to degraded quality of graphene in the grain boundaries, often yielding a mix of a-C and graphene. This leads to the small fraction of graphene observed at the FePt grain boundaries appearing to inhibit the formation of tall columnar grain structures in the FePt / graphene media. Low-rate sputter depositions of Corban are required to introduce a sufficient fraction of Graphene as GBM into the FePt media, which remains a significant challenge regarding FePt-graphene-based HAMR media. Achieving a higher graphene fraction in the grain boundaries could potentially enhance the grain isolation and enable the formation of taller, columnar FePt grains desirable for HAMR media applications.

**Acknowledgements**

This research was funded in part by the Data Storage Systems Center at Carnegie Mellon University and all its industrial sponsors and by the Kavcic-Moura Fund at Carnegie Mellon University. The authors acknowledge the use of the Materials Characterization Facility at Carnegie Mellon University supported by Grant No.MCF-677785.

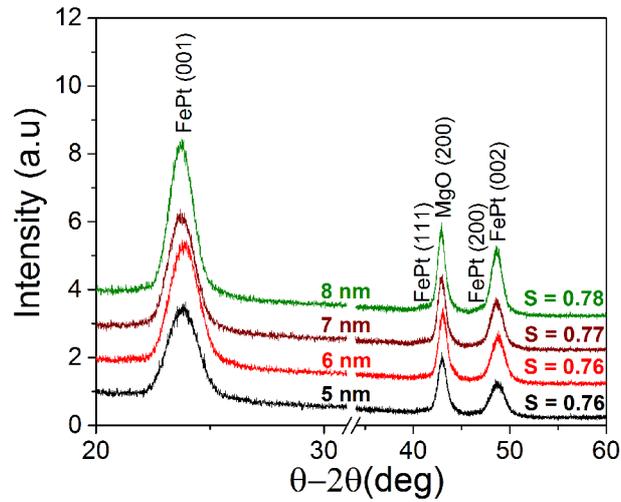

Figure.1 XRD patterns of the FePt-C-ML ($t$ = 5, 6, 7 and 8 nm) film stack deposited at 650°C. The order parameters ($S$) for these films with various thickness are labeled.

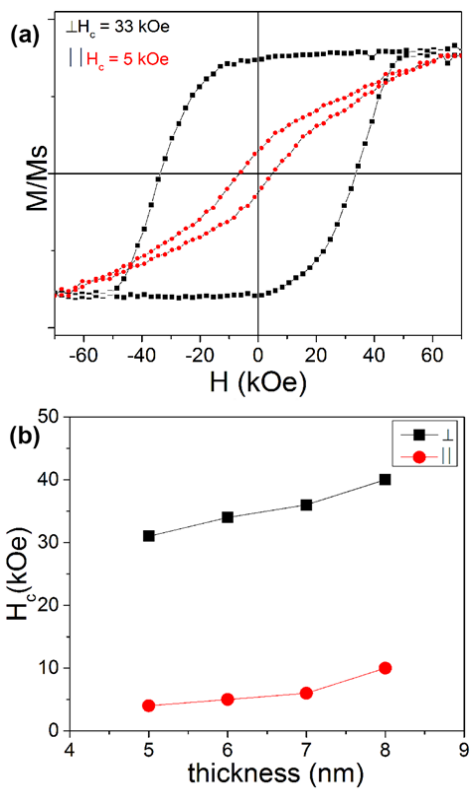

Figure.2 (a) Plot of normalized magnetization (M/Ms) vs. magnetic field (H) of the FePt-C-ML ($t$ = 5 nm) sample. (b) Perpendicular and in-plane $H_C$ of FePt-C-ML (t= 5, 6, 7, 8 nm) film stacks deposited at 650°C

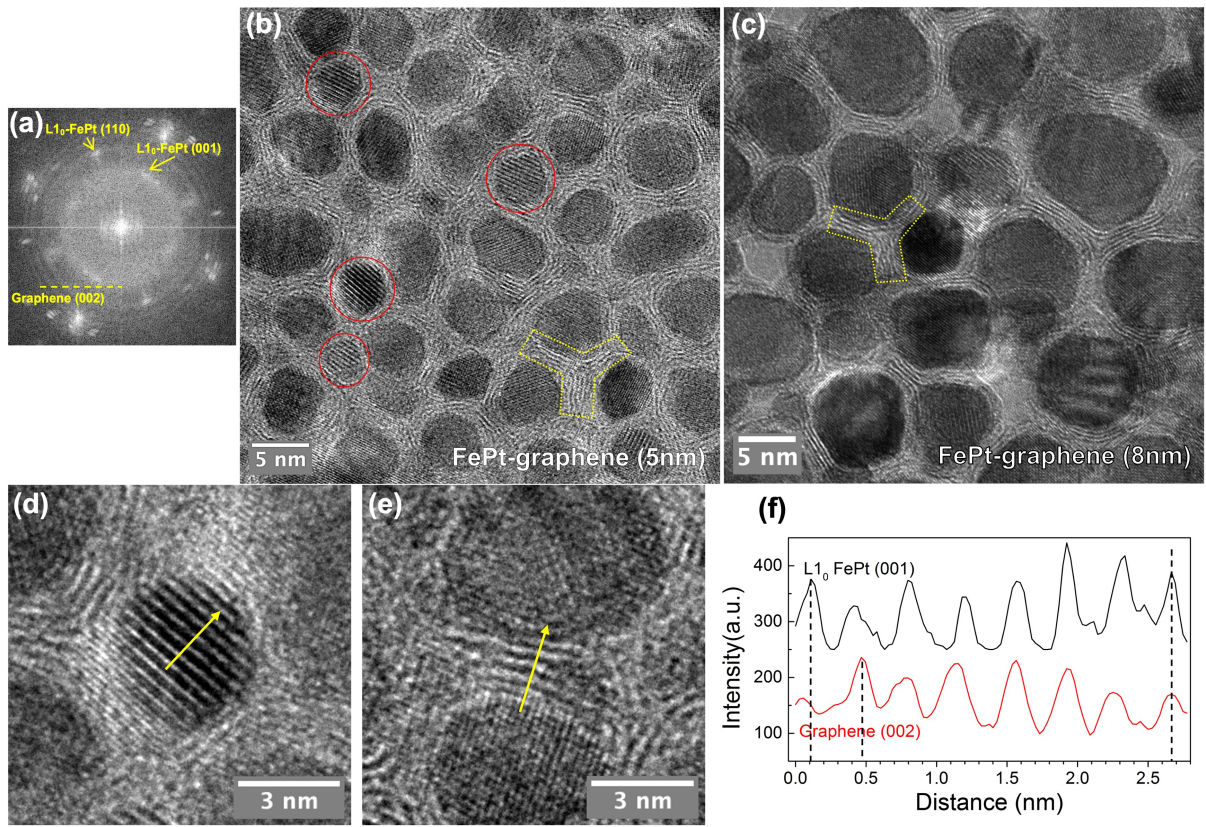

Figure.3 (a),(b) In-plane TEM micrograph of the 5-nm FePt-C(graphene)-ML film and its FFT pattern. Red circles label the $L1_0$-FePt grains with the c-axis in the plane. Yellow dotted frames label the Y-shape junctions of graphene multilayers. (c) In-plane TEM micrograph of the 8-nm FePt-C(graphene)-ML. (d),(e) Lattice spacing measurements on the 5-nm FePt-C(graphene)-ML and (f) the corresponding line intensity profiles.

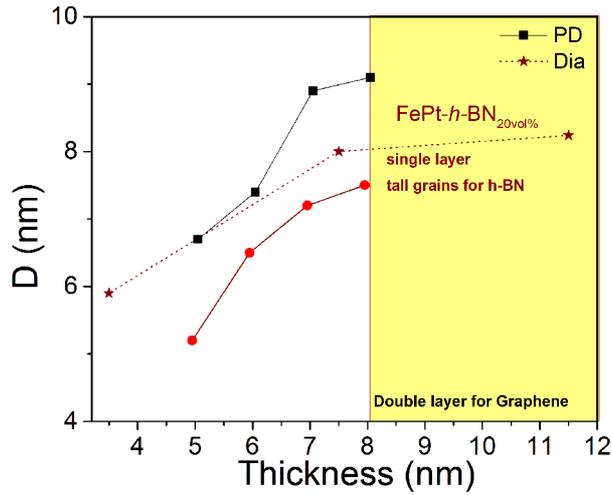

Figure.4 Grain diameter and pitch distance of FePt-C(graphene)$_{30vol\%}$ film for different thicknesses. Dotted line with star symbol represents the data of FePt-(h-BN)$_{20\ vol\%}$ media thin films.

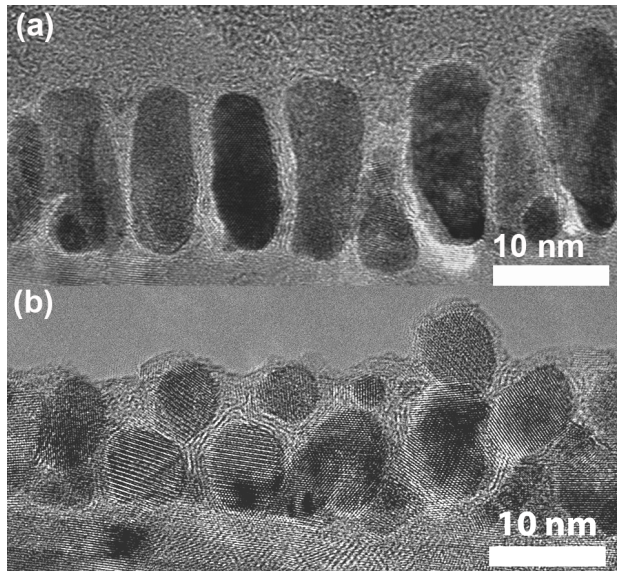

Figure.5 Cross section HRTEM images of (a) FePt-(h-BN)$_{26vol\%}$ film (deposited with substrate bias 3W) with thickness of 12 nm; (b) FePt-C(graphene)$_{30vol\%}$ film (deposited with substrate bias 5W) with thickness of 11 nm.